\documentstyle[prl,aps,psfig,epsf]{revtex}
\newcommand{\ave}[1]{\mbox{$\langle #1 \rangle$}}
\newcommand{\aave}[1]{\mbox{$\langle \langle #1 \rangle \rangle$}}

\newcommand{\beq}{\begin{equation}}
\newcommand{\eeq}{\end{equation}}
\newcommand{\beqa}{\begin{eqnarray}}
\newcommand{\eeqa}{\end{eqnarray}}
\newcommand{\bmath}{\begin{mathletters}}
\newcommand{\emath}{\end{mathletters}}
\newcommand{\nbar}{\mbox{$\bar{N}$}}
\newcommand{\rhobar}{\mbox{$\bar{\rho}$}}
\newcommand{\Phibar}{\mbox{$\bar{\Phi}$}}
\newcommand{\Psibar}{\mbox{$\bar{\Psi}$}}
\newcommand{\Ebar}{\mbox{$\bar{E}$}}
\newcommand{\Fbar}{\mbox{$\bar{F}$}}
\newcommand{\estyle}{\mbox{${\cal E}$}}
\newcommand{\estylebar}{\mbox{$\bar{{\cal E}}$}}
\newcommand{\sigmabar}{\mbox{$\bar{\sigma}$}}
\newcommand{\vtrapbar}{\mbox{$\bar{V}_{trap}$}}
\newcommand{\Omegabar}{\mbox{$\bar{\Omega}$}}
\newcommand{\xibar}{\mbox{$\bar{\xi}$}}
\newcommand{\rbar}{\mbox{$\bar{r}$}}
\newcommand{\Deltabar}{\mbox{$\bar{\Delta}$}}
\begin{document}
\draft
\preprint{ICTP-SISSA}
\advance\textheight by 0.2in
\twocolumn[\hsize\textwidth\columnwidth\hsize\csname@twocolumnfalse%
\endcsname 
\title{Macroscopic Angular Momentum States of Bose-Einstein Condensates\\ in
Toroidal Traps}

\author{M.~Benakli$^1$, S.~Raghavan$^1$, A.~Smerzi$^2$, S.~Fantoni$^{1,2}$,
and S.~R.~Shenoy$^1$}
\address{
$^1$ International Centre for Theoretical Physics,
 I-34100, Trieste, Italy \\
$^2$ International School for
 Advanced Studies and Istituto Nazionale de Fisica
della Materia,
I-34014, Trieste, Italy
}
\date{\today}
\maketitle
\begin{abstract}
We consider a Bose-Einstein condensate (BEC) 
of $N$ atoms of repulsive interaction $\sim
U_0$, in an elliptical trap, axially pierced by a Gaussian-intensity laser
beam, forming an effective (quasi-2D) toroidal trap with minimum at radial
distance $\rho = \rho_p$. The macroscopic angular momentum states
$\Psi_l(\rho,\theta) \sim \sqrt{N}\Phi_l(\rho) e^{i l \theta}$ for integer
$l$ spread up to $\rho \lesssim \rho_{max} \sim (NU_0)^{1/4} \gg
\rho_p$.  The
spreading lowers
rotational energies,
 so estimated low metastability
barriers 
 can support large $l \lesssim l_{max}  \sim (NU_0)^{1/4}, \lesssim 10$ for
 typical parameters. The $l$-dependent density
profile $|\Phi_l(\rho)|^2 - |\Phi_0(\rho)|^2$ is a signature of BEC rotation.
Results are insensitive to
off-axis laser displacements $\rho_0$, for $\rho_0/\rho_{max} \ll 1$.
\end{abstract}

]
%\section{}
%\twocolumn
Bose-Einstein condensates (BEC) of atoms in magnetic traps have been the
focus of intense recent activity
\cite{bec123,andrews,edwards,bay-pet,griffin,javan1,sfgs,rsfs,zapsoleg,%
dbec-vortex,str-vortex,rokh-vortex}.
 Coherence of the BEC wavefunction was demonstrated
 by formation of atomic interference fringes, on switch-off of
weakly-coupled traps \cite{andrews}. Non-destructive tests of phase-coherence,
through  the Josephson-like effects have been proposed
\cite{javan1,sfgs,rsfs,zapsoleg},
 including novel self-trapping phenomena
\cite{sfgs,rsfs}. Other phase-coherence signatures in trapped BEC would be
of much interest. 
\par
In bulk superfluids/superconductors, equilibrium quantized
 vortices \cite{wil-pack-ess-trau,vinen},
provided classic evidence of phase coherence. In multiply-connected
axial-hole 
geometries  (e.g. thick superconducting cylinders
\cite{little-parks} or superfluids in narrow rings),
  superflows
with wavefunction $\Phi_l(\theta) = |\Phi| e^{i l \theta}$, and $|\Phi|$
uniform \cite{fetter-earlyvortex,bloch-putter,rokh-vortex2-gold} maintain 
integer-$l$ metastable quantized
rotational states.
 BEC $l\!\neq \!0$ states in a He II like uniform-density limit have been
considered \cite{bloch-putter,rokh-vortex2-gold},
 with the BEC
rigidly restricted to flat-potential regions by (narrow-ring) rigid walls.
However, one
needs to go beyond this `square-well container' limit to capture the
characteristic $N$-dependent spread of BEC in a 
polynomial trap.
 The
Gross-Pitaevskii equation (GPE) \cite{gpe}
 for a non-uniform BEC includes
both (repulsive) atomic collisions that spread BEC,
 and trap potentials, that confine it.
GPE vortex solutions  centered in
 (simply-connected) harmonic traps have been found
\cite{dbec-vortex,str-vortex},
 but could be unstable to outward vortex displacements
\cite{rokh-vortex}. 
\par
The investigation of $N$-atom {\em non-uniform metastable} states, of
macroscopic angular momentum $N l \hbar$, in {\em multiply-connected}
toroidal traps are thus  important as possible signatures of BEC phase
coherence. Central theoretical issues include the existence
of such GPE states; the
mixing of $l$-states by nonlinear GPE terms; the effects of off-center
(azimuthally asymmetric) displacements of the toroidal hole; the existence
of metastability barriers, and the
decay-channel limiting of accessible $l$-values.
\par
\begin{figure}
\psfig{file=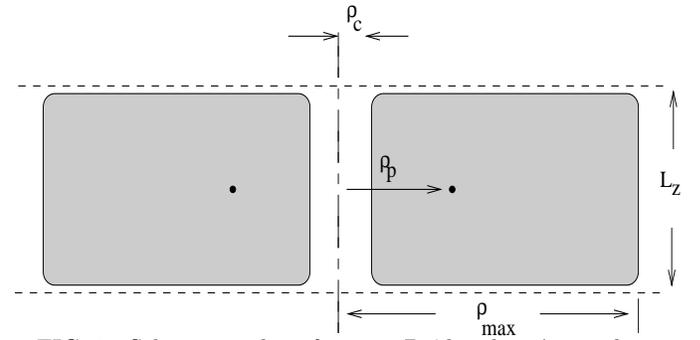,height=1.75in,width=3.5in}
\caption{Schematic plot of quasi-2D `doughnut'
 toroid geometry with laser along the
$z$ axis. 
The trap potential rises sharply
(dashed boundary) at $L_z$. Here 
$\rho_c,\rho_p$, and $\rho_{max}$ are the effective
core radius, toroidal axis, and extent of 
the shaded BEC cloud respectively.}
\label{fig:schematic}
\end{figure}
\begin{figure}
\psfig{file=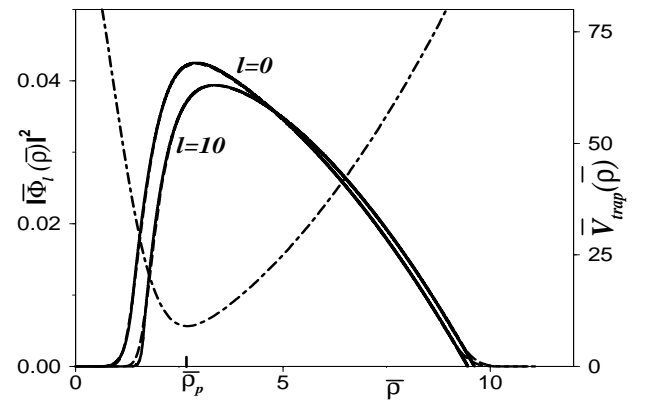,%
height=2.5in,width=3.5in}
\caption{Trap potential (right vertical axis, dash-dot line) 
$\bar{V}_{trap}(\rhobar)$ 
versus  cylindrical-coordinate
radius, $\rhobar$,(with bars
denoting scaling in harmonic trap energy/length).
 $\rhobar_p$ is
 the effective toroidal trap minimum.
Scaled BEC density
$|\Phibar_l(\rhobar)|^2$  (left vertical axis) versus 
$\rhobar$ from analytic (solid line) and 
numerical (dashed line) results. Parameters are as in text.}
\label{fig:wfn}
\end{figure}

In this Letter, we consider $N$ atoms in the BEC,
 of interatom scattering $\sim
U_0 >0$ in an elliptical trap, with an axial `hole' (or strong
potential barrier) drilled by an intense off-resonant 
laser beam, forming an effective
toroidal trap. The $T=0$ 
GPE is solved 
for
a particular quasi-2D toroidal trap,
with the Thomas-Fermi approximation (TFA)
\cite{edwards,bay-pet,griffin} outside an effective core, and a polynomial
solution inside it. 
 The cylindrical-coordinate
wavefunctions $\Psi_l(\rho,\theta,z) \sim \sqrt{N} \Phi_l(\rho)
e^{il\theta} $  are used as bases for
variational states.  
 A dimensionless
expansion 
parameter
$\delta$, that is small for strongly interacting and dense trapped BEC,
 appears naturally.
We find that the $l=0$ state $\Phi_{l=0}(\rho)$ of energy
$E_0 \sim \delta^{-1}$
peaks at the 
toroidal axis, spreading out to a radial distance $\rho = \rho_{max} \sim
\delta^{-1/2}$. The centrifugal force induces a density fall-off
 $|\Phi_l(\rho)|^2 - |\Phi_{l=0}(\rho)|^2 \sim l^2 \delta^3
[\ln(\delta^{-2}) -
\rho_{max}^2/\rho^2]$, that is a rotation signature
of the $l \neq 0$ BEC states, of rotational energy 
$E_l - E_0 \sim l^2 (\ln \delta)\delta \sim l^2/\rho_{max}^2$.
Off-axis displacements $\sim \rho_0$ of the toroidal `hole' are unimportant 
if $\rho_0/\rho_{max} \ll 1$.
 The self-interaction $U_0 |\Psi|^2$, rather than
destroying $l$-states, provides metastability barriers $E_B$ against 
$l\! \rightarrow\! l\!-\!1$ reductions. Other decay channels, by emission of
unit-vortices, or of Bogoliubov quasi-particles, have comparable thresholds,
$E_B \sim O(1)$. Increasing the BEC spread $\rho_{max}$ lowers the $l \neq
0$ cost, $E_l-E_0 \sim l^2/\rho_{max}^2$, so 
 $\sim O(1)$ barriers $E_B \gtrsim
E_l-E_0$ can  still maintain large $l$ values, $l \lesssim l_{max} \sim
\delta^{-1/2}$. These quasi-2D trap
 results, as well as a square-well container limit,
are recovered, from a general $d$-dimensional,
anharmonic-trap scaling analysis. We now outline arguments for the above
conclusions. 
\par
i) {\em GPE and $l$-states}:  The macroscopic condensate wavefunction
$\Psi(\rho,\theta,z) = \sqrt{N}\Phi(\rho,\theta,z)$ obeys the GPE in
cylindrical coordinates:
\beqa
\label{eq:gpe}
\left[-\frac{\hbar^2}{2m}\left(\frac{\partial^2}{\partial \rho^2} +
\frac{1}{\rho}\frac{\partial}{\partial \rho} +
\frac{1}{\rho^2}\frac{\partial^2}{\partial \theta^2} + 
\frac{\partial^2}{\partial z^2} \right) + \right. \nonumber \\
\left. V_{trap}(\rho,\theta) + 
V_{trap}(z) + U_0 |\Psi|^2\right]\Psi = E \Psi,
\eeqa
where $U_0=4\pi\hbar^2 a/m$ and $a \, (m)$ is the atomic scattering length
(mass) and $E$ is the (single-particle) energy.
$V_{trap}(\rho,\theta)(V_{trap}(z))$ is harmonic with curvature
  $\omega_{\|}(\omega_z$). However, $V_{trap}(z)$ rises sharply 
for $|z| \gtrsim L_z/2$, as in Fig.~\ref{fig:schematic}.
The interacting BEC cloud is blocked from expanding beyond 
 $|z| \lesssim L_z/2$, spreading,
  for 
$\omega_{\|}^2 \rho_{max}^2 \gg \omega_z^2 L_z^2$,
 only in the $\rho$-direction
 with increasing $N$.
The wavefunction is quasi-2D, uniform $\sim 1/\sqrt{L_z}$ in the $z$-direction.
 An off-resonant
 gaussian-profile laser barrier
of high intensity $ V_c$, and narrow width $ 2 \sigma$, directed along
the $z$-axis, yields an axially symmetric `doughnut' trap
 $V_{trap}(\rho,\theta) \rightarrow
V_{trap}(\rho)$, where
\beq
\label{eq:trap}
V_{trap}(\rho) = \frac{1}{2} m \omega_{\|}^2 \rho^2 + 
V_c \, e^{-\rho^2/2\sigma^2}.
\eeq
 The BEC
wavefunction is, with $l=0,1,2,...$, 
\beq
\label{eq:bec2dwfn}
\Phi_l(\rho,\theta,z) \rightarrow \Phi_l(\rho,\theta) =
\frac{\bar{\Phi}_l(\bar{\rho})}{\sqrt{L_z} r_{\|}}\frac{e^{i l
\theta}}{\sqrt{2 \pi}}.
\eeq
We  scale lengths (energies) in the $V_c=0$ harmonic trap length
$r_{\|}=(\hbar/m\omega_{\|})^{1/2}$ (energy = $\hbar \omega_{\|}/2$), so 
$\bar{\rho} \equiv \rho/r_{\|},\bar{\sigma}=\sigma/r_{\|},
\bar{E}_l = E_l/\frac{1}{2}\hbar\omega_{\|},\bar{V}_c\equiv
V_c/\frac{1}{2}\hbar\omega_{\|}$. Then  Eq.~(\ref{eq:gpe}), in dimensionless
form, is
\beqa
\label{eq:gpestd}
\left[-\left(\frac{d^2}{d\bar{\rho}^2} +
\frac{1}{\bar{\rho}}\frac{d}{d\rhobar}\right) + \frac{l^2}{\rhobar^2} +
\bar{V}_{trap}(\rhobar) + \right. \nonumber \\
\left. \nbar |\Phibar_l(\rhobar)|^2\right] \Phibar_l(\rhobar) =
\Ebar_l \Phibar_l(\rhobar),
\eeqa
with $\nbar \propto N$ defining a central dimensionless parameter,
\beq
\label{eq:defns}
\delta \equiv \nbar^{-1/2} \equiv (N_0/N)^{1/2}
\; ; \; N_0 \equiv L_z/4 a.
\eeq
As shown in Fig.~\ref{fig:wfn},
 the toroidal trap potential
$\bar{V}_{trap}(\rhobar) = \rhobar^2 + \bar{V}_c e^{-\rhobar^2/2\sigmabar^2}$
 of Eq.~(\ref{eq:gpestd}) 
has a minimum around the (circular)
toroidal axis at $\rhobar = \rhobar_p \equiv \sqrt{2} \sigmabar
[\ln(\bar{V_c}/2\sigmabar^2)]^{1/2}$ : $\bar{V}_{trap}(\rhobar) 
\approx
(\rhobar_p^2/\sigmabar^2)(\rhobar - \rhobar_p)^2 + \bar{V}_p$, where
$\bar{V}_p = \rhobar_p^2 + 2\sigmabar^2$. A non-interacting BEC in such a
minimum has a gaussian width $r_p = r_{\|}(\sigma/\rho_p)$, i.e., the 2D
`volume' occupied is $\sim r_{\|}^2$. The interacting BEC spreads (as seen
later) to $\rhobar_{max} \sim \nbar^{1/4}$, with a volume
$\rho_{max}^2 \gg r_{\|}^2$. Thus $\delta$ of Eq.~(\ref{eq:defns}) 
is essentially the BEC volume ratio,
 $\delta \sim r_{\|}^2/\rho_{max}^2 \ll 1.$ Note
that $\delta \sim 1$ 
 for the `square-well' container
 \cite{bloch-putter,rokh-vortex2-gold,zha-price}, where the BEC cannot
expand.
\par
We have solved Eq.~(\ref{eq:gpestd}) numerically. In the regime $\delta \ll
1$, and for a sufficiently low-width laser-hole, $\sigmabar \ll
\delta^{-2/13}$, we find analytic results can be obtained. 
Dropping derivatives $\sim \delta \ll 1$ in
Eq.~(\ref{eq:gpestd}), we have, for 
$\rhobar$ such that
$\rhobar_{max} > \rhobar > \rhobar_c >
\rhobar_{min}$, the TFA wavefunction
\beq
\label{eq:tfa1wfn}
\Phibar_l(\rhobar) = \left(\frac{\Ebar_0}{\nbar}\right)^{1/2}\left[ 
1 + \left[\estylebar_l - l^2/\rhobar^2 -
\vtrapbar(\rhobar)\right]/\Ebar_0\right]^{1/2},
\eeq
where $\estylebar_l \equiv \Ebar_l - \Ebar_0$ is the $l\neq0$ energy
per particle above the
ground state and $\Phibar_l(\rhobar_{max}) = \Phibar_l(\rhobar_{min}) = 0$. The
wavefunction is zero for $\rhobar > \rhobar_{max}$;
it is matched for $\rhobar < \rhobar_c$ to a polynomial solution
$\Phibar_l^<$ inside an effective `core' $\rhobar_c \gtrsim \rhobar_{min}$.
$\Phibar_l(\rhobar)$ peaks at $\rhobar = \rhobar_{peak} \approx \rhobar_p +
(\sigmabar^2/\rhobar_p^4) l^2$. The energies are determined by normalization,
\beq
\label{eq:tfa1normn}
\int_{\rhobar_c}^{\rhobar_{max}} d\rhobar \rhobar |\Phibar_l(\rhobar)|^2 + 
\int_0^{\rhobar_c} d\rhobar \rhobar |\Phibar_l^<(\rhobar)|^2 = 1.
\eeq
With the first integral dominating, we have to 
$O(\delta^2)$,
\beq
\label{eq:tfaenergy}
\Ebar_l = \Ebar_0 + \lambda l^2 \delta \; ;
 \; \Ebar_0 = 2
\nbar \delta = (4 \nbar)^{1/2}.
\eeq
 Here, $\lambda \equiv
\ln(\rhobar_{max}/\rhobar_c),\rhobar_c \sim
 \sigmabar [\ln(\bar{V}_c/\Ebar_0)]^{1/2}$,
 so the results are only very weakly dependent 
($\lambda \sim \ln \rhobar_c \sim \ln[\ln[\bar{V}_c/\sigmabar]]$) on the
toroidal core region and how it is modeled. The interaction pushes the BEC
far beyond the toroidal axis, to $\rhobar = \rhobar_{max}(l) \gg \rhobar_p$,
 and with
$\rhobar_{m} \equiv \rhobar_{max}(l=0)$, we have
\beq
\label{eq:rhomax}
\rhobar_{max}(l) = \rhobar_{m}(1+\frac{\lambda l^2 \delta^2}{4}) \; ; \;
\rhobar_{m} \simeq (4
\nbar)^{1/4} = \Ebar_0^{1/2}.
\eeq
The local density difference due to centrifugal effects $\sim
l^2/\rho^2$ in Eq.~(\ref{eq:gpestd}) is 
\beq
\label{eq:difflddiff}
|\Phibar_l(\rhobar)|^2 - |\Phibar_{l=0}(\rhobar)|^2 = \lambda l^2 \delta^3
[1 - (\rhobar_{m}^2/2\lambda\rhobar^2)],
\eeq
with $\Phibar_l$ crossing $\Phibar_0$ at $\rhobar \approx
\rhobar_{m}/\sqrt{2\lambda}$. 
\par
Returning to the $\rhobar<\rhobar_c$ series solution, $\Phibar_l^<(\rhobar)
= \sum_{i=0}^2 a_{2i} \rhobar^{2i+l} $, we find that  the 
coefficients are
$a_0 \sim a_4/\bar{V}_c^2, a_2\sim a_4/\bar{V}_c$, with $a_4 \sim
\sigmabar^{3/2}\delta$.
The TFA $\Phibar_l(\rhobar)$  falls off as $\sim (\rhobar -
\rhobar_{min})^{1/2}$, with $\rhobar_{min} \simeq
\sqrt{2}\sigmabar[\ln(\bar{V}_c/\Ebar_0)]^{1/2}$.
 The wavefunctions and derivatives can be
matched at a $\rhobar_c \equiv [(l+4)/(l+7/2)]\rhobar_{min} >
\rhobar_{min}$, and the  $\Phibar_l^<$ integral in
Eq.~(\ref{eq:tfa1normn}) is $\sim a_4^2 \rhobar_c^{10} \sim \sigmabar^{13}
\ln(\bar{V}_c \delta) \delta^2 \ll 1.$ 
\par
Figure \ref{fig:wfn} shows $|\Phibar_l(\rhobar)|^2$ versus $\rhobar$,
 for $l=0$ and $10$, for parameters as
given later. The analytic approximation (solid line) closely matches the 
(dashed line) numerical solution of Eq.~(\ref{eq:gpestd}).
The centrifugal 
reduction and shift of the density peak   ($\sim 10\%$ for
$l=10$), is  a
rotation signature, possibly detectable by phase contrast imaging
\cite{bec123}. 
 The super-current density is $j \sim  \hbar/(i m 2\pi \rhobar)
\Psi^\ast \frac{\partial}{\partial \theta} \Psi =
N |\Phibar_l(\rhobar)|^2 v_{s\theta}$, where
the azimuthal velocity
 $v_{s\theta} = (\hbar
l/2 \pi m\rho)$, and $j$ vanishes at the origin as 
$j \sim \rho^{2l+3}$. 
The laser hole `pins' the azimuthal velocity of 
average angular momentum 
 $\ave{\Psi_l|(\hbar/i)\partial/\partial\theta|\Psi_l} = Nl\hbar$,
  suppressing displacement instabilities
\cite{rokh-vortex}. 
\par
%\item
ii)  {\em Metastability barriers}: The
 nonlinear term in the GPE
could cause mixing of  $l$-states
 and  $l \! \rightarrow \! l\!-\!1$ decays, but
instead, induces barriers. 
 A GPE trial function for full superposition is
 $\Psibar(\rhobar,\theta)/\sqrt{N} = [C_l\Phibar_l(\rhobar,\theta) +
C_{l-1}\Phibar_{l-1}(\rhobar,\theta)]$, with 
 variational constants, $|C_l|^2 + |C_{l-1}|^2
= 1$. With an external angular velocity 
$\Omega \equiv \Omegabar \omega_{\|}/2$ in the hamiltonian
 $\hat{H} \rightarrow \hat{H} - \hbar \Omega \hat{L}$,
 the free energy
 functional $\Fbar(\Psibar)$ (whose variation is Eq.~(\ref{eq:gpe})) is 
given by,
\beqa
\label{eq:vary}
\frac{\Fbar}{N} &=& \sum_{p=l,l-1} \{\Ebar_{p}(\Omegabar) - I_{p,p}
(1-|C_p|^2)\}|C_p|^2  \nonumber \\
&+&  4 |C_l|^2 |C_{l-1}|^2 I_{l,l-1},
\eeqa
where $I_{l,l'} \equiv \frac{1}{2}\nbar \aave{
|\Phibar_l(\rhobar)|^2 |\Phibar_{l'}(\rhobar)|^2}$, with
$\aave{...} \equiv \int_0^\infty d\rhobar \rhobar^{d-1} ... $, and
$d=2$ at present, and
 $\Ebar_l(\Omegabar) \equiv
\Ebar_l - l\Omegabar$. 
 The $l$-state $(C_l,C_{l-1})=(1,0)$
  is separated from the $l\!-\!1$-state
$(0,1)$   by a
barrier of height
 $\Fbar_B/N \equiv 
\Ebar_B \equiv \Ebar - \Ebar_l(\Omegabar)$, larger than the
splitting,
$\Delta \Ebar_l = (2 l - 1) \lambda \delta$. The `full-overlap' barrier
$\Ebar_B = \Ebar_{B1}$ is 
at $|C_l|^2=|C_{l-1}|^2 = \frac{1}{2}+O(\delta^2)$,
\beq
\label{eq:vary_barrier}
\Ebar_{B1} =  2\Ebar_0[1 - \lambda(l^2 +
(l-1)^2) \delta^2] 
- (\Delta \Ebar_l - \bar{\Omega}).
\eeq
Thus $\Ebar_{B1} \sim \Ebar_0
\sim \nbar^{1/2}$,
and the $l$-state $(1,0)$ 
 is
preserved by metastable barriers for 
$\estylebar_l = l^2\lambda/\sqrt{\nbar} < \Ebar_{B1}$ or $l<
\nbar^{1/2}$. These barriers arise 
because interatomic
collisions disfavor the transition-state 
slowing down of a minority fraction ($<1/2$) of the
rotating flow. 
\par
We now consider a superposition of 
non-uniform $l$-states, overlapping only at an interface, lowering the
  $l,l-1$ 
interspecies scattering. With $\nu_l,\nu_{l-1}$, the
fractional volumes occupied, and 
$\nu_0 \equiv (\nu_l+\nu_{l-1}-1) > 0$ the overlap fraction,
we take $\Phibar_p(\rho) \rightarrow
\Phibar_p(\rhobar)g_p (\rhobar)$ where the step-function
$g_{l-1}(\rhobar) \, (g_l(\rhobar))$ is unity for 
$\nu_{l-1}\rhobar_m^2 > \rhobar^2 > \rhobar_c^2$
 (for 
$\rhobar_{m}^2 > \rhobar^2 > (1-\nu_l) \rhobar_{m}^2) $ and zero 
otherwise. Repeating the previous argument with these wavefunctions, we find
that 
 the intra-species
scattering is increased by the 
confinement to a reduced volume $\nu_p
\rhobar_{m}^2 < \rhobar_{m}^2$, so the TFA energies are raised, 
$\Ebar_p(\nu) \simeq I_{p,p}(\nu)  \simeq \Ebar_p(\nu=1)/\nu_p$.
 The inter-species scattering gives 
$I_{l,l-1}(\nu)  \simeq
\nu_0 \nbar^{1/2}/\nu_l \nu_{l-1}$. 
The lowest interface-overlap barrier $\Ebar_{B2}$ obtains for approximately
equal volume fractions $\nu_l=\nu_{l-1}\gtrsim 1/2$, and for an annular
overlap volume of radius $\rhobar_m/\sqrt{2}$ and thickness $\sim \xibar$.
Thus $\nu_0 = \sqrt{2}\xibar\rhobar_m$; $\Ebar_{B2}$ is a fraction of the
full-overlap barrier, $\Ebar_{B2} \sim
(\sqrt{2}\xibar/\rhobar_m)\Ebar_{B1}$. 
(Gradient contributions omitted in TFA would be comparable). Here
$\xibar \sim 1/[\nbar |\Phibar_l(\rhobar_p)|^2]^{1/2} \sim 1/\Ebar_0^{1/2}$
 is a 
`healing length', so $\Ebar_{B2} = 2\sqrt{2} \sim O(1)$,
 maintaining $l$-states for 
$l \lesssim  l_{max2} = (2\sqrt{2}/\lambda)^{1/2}\delta^{-1/2},
 \sim \nbar^{1/4}$.
The external angular velocity induces an 
$l \rightarrow l-1$ transition 
only above critical values, $|\bar{\Omega}| \geq
|\bar{\Omega}_{cl}| = |(\bar{E}_B - \Delta \bar{E}_l)|$. This 
$\bar{\Omega}_{cl} \sim \Ebar_{B2}$ estimate exceeds an 
$\Omegabar_{cl} \sim |\Delta \Ebar_l|$ estimate \cite{bay-pet}
based on the splitting alone. 
\par
Wedge-shaped $l\!-\!1$ regions, expanding
azimuthally, would have $\sim (2\xibar/\rhobar_{m})\Ebar_{B1}$ 
interface costs,  and similar $l_{max} \sim \nbar^{1/4}$ scaling.
 2D phase-slip saddle-point solutions  \cite{rokh-vortex2-gold,zha-price} 
would
also provide
such
 barriers.
\par
%\item
iii) {\em Off-center toroidal hole}: In practice, the toroidal `hole' of
Fig.~(\ref{fig:schematic}) could be drilled off the elliptical-trap axis, or
the laser could fluctuate in profile and position. With
 a displacement $\vec{\rho} =
(\rho_0,\theta)$ of the $\sim \rho^2$ elliptical trap,  $V_{trap}(\rho)
\rightarrow V_{trap} + \rhobar^2 - 2\rhobar_0 \rhobar \cos \theta$,
modifying  the TFA
solution of Eq.~(\ref{eq:tfa1wfn}). 
 Repeating the above argument,  the BEC cloud becomes
anisotropic, extending to $\rhobar_{m} = \rhobar_0 \cos \theta +
\Ebar_0^{1/2}$, but
 global averages of the  energy/angular momentum are
unaffected, to order  $\sim \rhobar_0/\rhobar_{m} \ll 1$.
 Although 
 $C_l C_{l-1}^3$ terms (previously zero by $e^{i l \theta}$ orthogonality) 
now enter in Eq.~(\ref{eq:vary}),
the barriers $\Ebar_{B1,2}$ to $l$-decay are similarly insensitive. 
\par
%\item
iv)  {\em $l$-decay channels}:
 An $l$-state could
decay through (rectangular) 
unit-vortex loops, with straight
segments of vorticity $J_z = -1$ reducing core 
vorticity  by
$l \rightarrow l-1$, that pushes  the $J_z = 1$ segment outwards by
a repulsive energy $\sim -(l-1) \ln \rho$.
 With  $|\Phibar(\rhobar_c)|^2 \propto
(\rhobar_c-\rhobar_m) \sim \rhobar_c \sim \sigmabar$, the nucleation cost
near the core is
 $\Ebar_{B3} \sim \xibar^2 \nbar |\Phibar_l(\rhobar_c)|^2 \sim \sigmabar$. 
Then  $\estylebar_l = \lambda l^2 \delta < \Ebar_{B3}$
accesses this decay channel
 only
 for $l \geq l_{max3} \sim
\sqrt{\sigmabar} \nbar^{1/4}/\sqrt{\lambda}$.
  (As $v_{s \theta} \rhobar \sim
\hbar l/2\pi m$ is a constant, $v_{s\theta}$ should not induce
a loop-expanding
 current-drive 
force \cite{vinen}.) 
\par
Another possible decay channel is a successive reduction $N \rightarrow N-2$
of BEC   $l$-state atoms, 
producing surface pairs of
 dissipative quasiparticles \cite{str-vortex} 
of Bogoliubov
excitation energy $2\Deltabar$, and angular momentum $l\pm q$ 
 \cite{bloch-putter,rokh-vortex2-gold}, that slow down to 
fall into the $l=0$ ground state. A detailed analysis 
\cite{fetter-earlyvortex} for toroidal traps is beyond the scope of this
paper; however, for $\estylebar_l \lesssim
 \Deltabar$, the
quasiparticle channel is not accessed for
 $l \lesssim \sqrt{ \Deltabar}\nbar^{1/4}$. In other contexts
\cite{collect-qp}, quasiparticle damping is small.
\par
%\item
v)  {\em Numerical estimates}: 
Parameters chosen are 
 $\omega_{\|} = 132 {\rm rad \, s}^{-1},
(1 nK), r_{\|} = 6.3 \mu m,
 L_z = 25 \mu m, 2\sigma = 12 \mu m,
a = 50 {\mbox \AA}, V_c = 63 nK$, and $m = 2\times 10^{-26}$kg
 ($\sim {\rm sodium}$).
Then  $\rho_p = 17.5 \mu m,N_0 \approx 10^3$, and for 
 $N = 10^6$, we have
$\delta = 1/\sqrt{\nbar} = 0.03.$ Physical magnitudes 
are then $E_0 = 32 nK, \lambda = 2$,
 $\estyle_l  = 0.031 l^2 nK, \rho_{m}  = 50 \mu m,
 \xi = 0.8 \mu m, v_{s \theta}(\rho) \leq v_{s
\theta}(\rho_c)  = 0.12 l {\rm mm \, s}^{-1},
\Omega_{cl} = $ 30 Hz,
and $l_{max2} \sim 10$. 
  Thermal excitations activated over $N E_B$ free energy barriers
are exponentially dilute, at low temperatures.
\par
%\end{enumerate}
Finally, we present a general $\nbar$-scaling argument.
 Consider a $d$-dimensional 
toroidal trap that, for distances $\rbar$ well away from the core, is
 $V_{trap} \sim \hbar \omega_{t}\bar{r}^\alpha$, with
$\bar{r} \equiv r/r_{t}$ and $r_{t} = (\hbar/m\omega_{t})^{1/2}$.
For $\alpha=2$ the trap is harmonic; for $\alpha \rightarrow \infty$, a
square-well container results, $V_{trap}=0(\infty)$ for $r<r_t(r>r_t)$. 
Then defining $\nbar \equiv (NU_0/\hbar\omega_t) r_t^d,
\Phibar(\vec{\rbar}) \equiv \Phi(\vec{r})/r_t^{d/2}, 
\Ebar_l \equiv E_l/\hbar\omega_t$, the TFA wavefunction,
$\Phibar_l = ([\Ebar_l-\rbar^\alpha-(l^2/\rbar^2)]/\nbar)^{1/2}$ is
almost flat, $|\Phibar_l(\rbar)|\simeq |\Phibar_0| \approx
(\Ebar_0/\nbar)^{1/2}$, with BEC spreading
 until $\rbar \simeq \rbar_m$,
when the confining potential rises to the energy, $\Ebar_l \sim
\rbar_m^\alpha$. With normalization $\aave{|\Phibar_l(\rbar)|^2} \simeq
|\Phibar_0|^2 \rbar_m^d = 1$, we obtain $\rbar_m = \nbar^{1/(\alpha+d)}$. Then
 $\estylebar_l \sim
l^2/\rbar_m^2$, and  the healing length is $\xibar \sim
(\nbar|\Phibar_0|^2)^{-1/2} \sim  \nbar^{-\alpha/2(\alpha+d)}$. The integral
 $I_{l,l-1} \sim \nbar\aave{|\Phibar_l(\rbar)|^2|\Phibar_{l-1}(\rbar)|^2}$ 
defines a full-overlap metastability barrier
$\Ebar_{B1} \sim I_{l,l-1} \sim (\nbar/\rbar_m^d)$;
 while for a $d-1$ dimensional $l,l-1$
interface of thickness $\xibar,\Fbar_{B2}/N \equiv 
\Ebar_{B2} =  (\xibar/\rbar_m)\Ebar_{B1}$.
 The vortex nucleation 
threshold is $\Ebar_{B3} \sim \xibar^2 \rbar_m^{d-2} \nbar |\Phibar|^2 \sim
\rbar_m^{(d-2)}$. The $l$-state energies $\estylebar_l $
exceed these barriers only for $l$ greater than
$l_{max1} \sim \nbar^{(2+\alpha)/2(\alpha+d)},
l_{max2} \sim \nbar^{(2+\alpha)/4(\alpha+d)},
l_{max3} \sim \nbar^{d/2(\alpha+d)}.$ For $\alpha=d=2$, we recover our
results, with $l_{max2}\sim l_{max3} \sim \nbar^{1/4}$.
  For $\alpha \rightarrow \infty,\rbar_m=1$,
 i.e.  a restricted $r_m = r_t$, and  the familiar
 He II healing length $\xibar \sim (N U_0)^{-1/2}$ is recovered, with
 energies $\Ebar_l \sim \nbar+l^2$, and barriers $\Ebar_{B1} \sim
\nbar,\Ebar_{B2}\sim\nbar^{1/2}, \Ebar_{B3} \sim O(1)$. 
The $l$-bounds are
$l_{max1} = \nbar^{1/2},l_{max2} \sim \nbar^{1/4},\l_{max3} \sim 0(1)$:
narrow-ring containers \cite{bloch-putter,rokh-vortex2-gold,zha-price}
would 
suppress vortex-loops and make quasi-1D phase-slips \cite{zha-price}
the lowest decay thresholds, $F_{B2} \sim N^{3/2} U_0^{1/2}$.
 Barriers  vanish in the ideal
gas limit, $U_0 \rightarrow 0$. 
\par
Experimental preparation of $l\neq0$ states could be through stirring normal
states  by a laser `paddle-wheel' \cite{bec123} and cooling below
transition; or by  rotating the trap \cite{trap-rotate} with
 $\Omega > \Omega_{cl}$, below $T_c$.
Experimental detection might be through the $l \neq 0$ density profiles, or
the Sagnac effect \cite{rokh-vortex2-gold} in large
azimuthal-velocity regions around the core. 
Another interesting setup is a 
 double-toroidal trap, with
 $N_1=N,l_1 =1 (N_2=N,l_2=0)$
for toroid $1(2)$, where the toroids initially are
separated by a blocking 
laser sheet, whose removal could
result in inter-trap 
`macroscopic quantum coherence' \cite{leggett}.
\par
In conclusion,  $l \neq 0$ macroscopic angular momentum
states of condensates, with $l$-dependent density profiles,
 can exist in toroidal traps,
maintained by atomic collisions,
 and robust to toroidal hole displacements, and to estimated decays,
 for $l$ not
too large.
 The observation of these states would be a signature of
macroscopic phase coherence of trapped Bose-Einstein condensates.
%\bibliography{/home/srirag/tex/bibtexbase/thesbib,%
%/home/srirag/tex/bibtexbase/bec,%
%/home/srirag/tex/bibtexbase/mesoscopic}

\end{document}